# Extending Social Resource Exchange to Events of Abundance and Sufficiency

Jonas Bååth*, Dept. of Service Management and Service Studies, Lund University, Sweden

Adel Daoud, Center for Population and Development StudiesHarvard T.H. Chan School of Public Health, Harvard University, The Unites States, and Dept. of Sociology and Work Science, Universityof Gothenburg, Swedem

*Corresponding author: jonas.baath@ism.lu.se

**Introduction**
This chapter identifies how scarcity, abundance, and sufficiency influence exchange behavior. Analyzing the mechanisms governing people's exchange of resources constitutes the foundation of several social-science perspectives (Baumgärtner et al., 2006; Daoud, 2018a, 2018b; Mullainathan & Shafir, 2013; Panayotakis, 2011; Turner & Rojek, 2001; Xenos, 2017). Neoclassical economics provides one of the most well-known perspectives of how rational individuals allocate and exchange resources. Using Rational Choice Theory (RCT), neoclassical economics assumes that exchange between two individuals will occur when resources are scarce and that these individuals interact rationally to satisfy their requirements (i.e., preferences). Scarcity is a state where available resources are insufficient to satisfy a set of requirements. This experience of insufficiency is arguably the primary motivator for why an individual engages in exchange. While RCT is useful to characterize interaction in closed and stylized systems, it proves insufficient to capture social and psychological reality where culture, emotions, and habits play an integral part in resource exchange (Archer & Tritter, 2000; Boudon, 2003; Hedström & Stern, 2008; Kahneman, 2011). Another deficiency of neoclassical economics is that needs, wants, and requirements are taken as a given. As scarcity is postulated, neoclassical economics cannot explain how scarcity may arise and justify how abundance and sufficiency are not possible motivators of human exchange.

Social Resource Theory (SRT) improves on RCT in several respects by making the *social nature of resources* the object of study (Foa & Foa, 1974). SRT shows how human interaction is driven by an array of psychological mechanisms — from emotions to heuristics. Thus, SRT provides a more realistic foundation for analyzing and explaining social exchange than the stylized instrumental rationality of RCT. SRT does not shy away from unpacking the inherently psychological and sociological concept of preferences by showing how and why scarcity motivates human action. However, while scarcity is undoubtedly a key motivator, SRT has no clear place for events of abundance and sufficiency as additional motivations to exchange resources. For example, as the scarcity of food or love may motivate individuals to overcome their predicament, an abundance of weapons or money may motivate individuals in a different way (Bååth, 2018; Daoud, 2018a; Dugger & Peach, 2009). Or, an abundance of food or books may have unintended consequences leading to starvation amid obesity (Daoud, 2007), or "paralysis due to information overflow" (Abbott, 2014). Likewise, sufficiency — neither too much nor too little — is the desired state of being for individuals attaining voluntary simplicity (Daoud, 2011b; Osikominu & Bocken, 2020). Consequently, if sufficiency and abundance affect human behavior besides scarcity, then how scholars should identify the dynamics among the three states that resources can attain: scarcity, abundance, and sufficiency.

The aim of our chapter is to synthesize and formalize a foundation for SRT using not only scarcity but also abundance and sufficiency. We will achieve this using mainly the theory of scarcity, abundance, and sufficiency (SAS) (Daoud, 2011a, 2018b). This theory focuses on explaining SAS rather than assuming them. Our SRT-SAS synthesis provides a general theory that enables formulating new predictions, hypotheses, and explanations about the dynamics among all three SAS events. Unlike neoclassical economics that mathematically and conceptually requires scarcity to function, SRT embraces a view of human actions that are compatible with scarcity and abundance and sufficiency.

**Background**

The assumption of scarcity is omnipresent in the social sciences. The most well-known example is neoclassical economics—the central tradition of economics—which assumes scarcity as the cause of all significant economic problems (Bronfenbrenner, 1962; Menger, 2007; Robbins, 2007; Zinam, 1982). For instance, Milton Friedman framed an economic problem in the following way: "An economic problem exists whenever scarce means are used to satisfy alternative ends. If the means are not scarce, there is no problem at all; there is Nirvana" (Friedman, 2009, p. 6). Following Friedman's argument, there would be no reason for markets to exist in a world of resource abundance. Neoclassical economics generalizes these sorts of economic problems to any human behavior by postulating that scarcity is both natural and universal (Becker, 1965) — it exists and perpetuates all behavior (Daoud, 2011a). RCT is the vehicle through which this sort of economic problem is generalized to any domain (Elster, 2015).

While SRT improves several aspects of how RCT depicts human behavior, we argue that SRT still assumes scarcity as the leading cause of human exchange. SRT is a theory about the configuration of societal structures and how their cognitive representations affect the dynamics of social resources exchange. SRT's definition of resources is inherently social, that is, a resource is a thing of value that is "transmitted through interpersonal behavior"(Foa & Foa, 1974, pp. 35–40). By emphasizing the social nature of resources, SRT defines six resource classes: love, status, information, money, goods, service. According to their relative position on two axes, these resource classes are positioned in a cognitive space: the degree of concreteness and particularity. Love is the most particular resource class since it is of great importance from whom one receives love. Money is the least particular resource since its value is less dependent on who gives and receives in than the other resource classes. The resulting four resource classes are in-between money and love on this axis. However, goods and services are the most concrete resource classes since they are the most tangible ones.

Conversely, information and status are the least concrete ones, due to their intangible and symbolic nature. Money and love occupy the middle-ground on the concreteness-axis (Foa & Foa, 1974, pp. 140–143). SRT mobilizes social-psychological explanations for how and why individuals exchange these resource classes, and what kind of distributional outcomes this exchange leads to (Foa & Foa, 1974, pp. 178–180). The following sections outline SRT.

*SRT and the assumption of scarcity*

SRT offers a sophisticated theory of how and why individuals exchange resources. We argue that its primary strength is that it emphasizes the social character of resources as exchange objects with a limited set of discrete qualities. This strength makes SRT a more realistic approach to exchange than neoclassical economics (which reduces all values to prices, losing out on the specific characteristics of different resources). However, we argue that SRT would benefit from developed theorizing in two regards: (A) how abundance and sufficiency affect the exchange and (B) the relation between different levels of social reality, primarily between the individual and the (societal) system levels. Developing these two areas, we show how the theory of SAS offers useful concepts for developing SRT.

The implicit assumption of scarcity in SRT is best shown by considering how Uriel G. Foa and Edna B. Foa define the concept need as "a state of deficiency in a given resource" for the individual (Foa & Foa, 1974, p. 130). To need a resource is to possess less than optimal amounts of that resource. That is scarcity. The opposite of needs is power, implying that the individual possesses more resources than she needs, and can use some resources for alternative ends than satisfying her needs (Foa & Foa, 1974, pp. 128–135). Power includes sufficiency and non-problematic forms of abundance. This distinction might seem like a simple binary of need and non-need. However, resource optimum is defined as a range, implying that exceeding said range would be suboptimal. In that event, the number of resources would amount to a negative abundance, causing the individual to experience some deficit or needs for welfare as a consequence (Foa & Foa, 1974, pp. 128–129). Jerald Greenberg (1981) develops the concept of needs by suggesting a distinction between objective and felt needs. Objective needs imply actual dearth.

Conversely, felt needs imply the impression of needs an individual might experience despite objectively possessing enough resources. Needs thus encompass all kinds of meaningful requirements for resources that an individual can experience. While Greenberg's definition implies that needs arise from scarce resources, it begs the question of how needs affect an individual's motivations to engage in exchange (motivations, in short).

The concept of optimal range defines the relationship between needs and motivations. This concept means that when an individual possesses an amount of a resource that falls within a specific range's limits, she experiences no needs (Foa & Foa, 1974, pp. 125, 128). This limit varies between individuals. This concept assumes that when an individual's available resources fall below this limit, scarcity arises and causes "motivational arousal" (Foa & Foa, 1974, p. 129). This arousal will result in two types of motivations: to maintain resource optimum or to maintain cognitive structures (Foa & Foa, 1974, p. 126). This chapter focus on the former motivation (Foa & Foa, 1974, pp. 5–13, 140–142).

The motivation to maintain resource optimum follows from the assumption that the individual experiences scarce resources as needs. Needs motivate individuals to engage in exchange to achieve resource optimum. Alternatively, the individual might fear a future scarcity, or, as Greenberg suggests, feel that she needs more of a resource despite objectively possessing sufficient or abundant amounts. In the end, the experience of needs becomes the motivation to exchange. Consequently, scarcity is essential for exchange behavior in SRT. However, an assumption of scarcity also implies that sufficiency and abundance would be comparably less developed.

While needs are important for explaining human action in general, SRT is diffuse about the motivations of actors who lack needs—those experiencing sufficiency or abundance. While resource optimum equals power, it remains unclear how power motivates individuals to engage in further exchange. One explanation is that individuals experiencing resource optimum are motivated to maintain their cognitive structures, meaning to avoid experiences that dissonate with the individual's imagined structures of social life. However, this explanation would demand that all individuals who exercise power do so to avoid cognitive dissonance. Such an argument is problematic, as it excludes the possibility that resource states might motivate individuals out of need (cf. Foa & Foa, 1974, pp. 140–141). For example, power could motivate the individual to accumulate more power, or to reduce the power of her competitors.

Resource optimum implies a range defined by an upper and lower limit for a given resource class (Foa & Foa, 1974, p. 128). An individual experiencing more than optimal resources would experience an emergence of a new need; a need for welfare due to the discomfort and suffering caused by "oversatitaion" (Foa & Foa, 1974, pp. 127–128, 138). An abundance of resources would — just like scarcity — arouse the motivation to maintain resource optimum. This consideration implies a definition of abundance that rests on the emergence of new scarcities. However, such an interpretation of abundance limits it to negative forms of abundance, meaning the events in which abundance results in different negative consequences. Any positive forms of abundance are left undefined.

By considering the relation between SRT's needs and motivations, our argument identifies that scarcity is intimately tied to how resource states affect exchange behavior. Nonetheless, SRT assumes this form of scarcity because it elevates scarcity to a more critical state than sufficiency or abundance. This elevation is evident in SRT's conception of sufficiency and abundance. The former is roughly the absence of need, and the latter is when new scarcities emerge due to super-optimal amounts of resources.

SRT's definition of scarcity and needs produces further conceptual tensions. Because of this overlap, SRT presupposes the assumption that all needs are fundamental and require satisfaction. Needs draw a normative boundary between the requirements that might be considered needs, and others which might be considered, e.g., wants or demands. This normative grounding is explicated in Greenberg's (1981) distinction between actual and felt needs, suggesting that the latter are less urgent or real. Such felt needs imply that individuals experiencing it would not objectively be in need. While Greenberg's argument opens up for the possibility of motivations among actors out of objective need, it still assumes that the motivation must be tied to resource deficiency. Beyond affirming the conclusion that SRT rests on an assumption of scarcity, the concept "need" further limits that assumption to a normative conception of how scarcity is experienced, because any other — felt or otherwise — the experience of scarcity would not be a legitimate requirement.

The scarcity assumption creates at least two problems for SRT's ambition to explain social resource exchange. The first problem is that the assumption of scarcity obscures how states of sufficiency and abundance might influence exchange in empirical work. Beyond being a normative concept, needs are hard to observe empirically because they demand a method that can distinguish needs from non-needs (Springborg, 1981). This is especially the case in measuring needs consistently, especially across different societies, cultures, and points in time. Kjell Y. Törnblom and Björn O. Nilsson (2008) engaged with this issue, seeking to nuance the conception of needs by qualifying different needs according to their importance. In another study, Törnblom and colleagues (2008) have explored if a resource's source might affect its satisfactory effect towards a need. However, these studies still raise questions regarding what extent satisfaction differs between needs other kinds of requirements, and how the relative importance of needs interact with other kinds of requirements

The second problem pertains to defining need as "a state of deficiency in a given resource." This definition raises questions about the context in which that deficiency occurs. Generally, SRT-studies limit their scope to the individuals involved in the exchange. Greenberg (1981) addresses this issue from a methodological angle, suggesting that SRT-based experiments on resource exchange do not account for the research subject's cultural context. However, Foa and Foa explicitly frame exchanges in institutions that define the "proper," i.e. legitimate, settings and relations between resources in exchange (1974, pp. 150–152). While we agree that the institutional setting is essential for analyzing resource exchange, Foa and Foa's definition leaves out what resources and exchange offers are available for exchange beyond the direct interaction.

In sum, SRT lacks an operationalization and theoretical clarification of how structural or systemic forces, such as institutions, constrain, and enable resource exchange. This issue stems from the assumption of scarcity. Because SAS events are experienced by individuals and groups embedded in social systems, a fully-fledged SRT theory has to account for SAS dynamics. While this focus on scarcity is purposeful, it allows SRT scholars to engage with some scientific issues; it leaves another set of relevant scientific problems unexplored. In contrast to neoclassical economics, SRT scope of analysis does not necessarily preclude the two other SAS events; however, to allow scholars to use these two concepts in tandem with scarcity, we need to redefine a set of concepts.

**On the dynamics of scarcity, abundance, and sufficiency (SAS)**

The SAS-theory is a socio-economic approach starting from the assumption that scarcity and abundance and sufficiency are important resource states for human interaction (Daoud, 2018b). As individuals orient themselves to these three states simultaneously, individuals' perception and interaction create dynamics in resource exchange. While assuming one event's existence over the other two is occasionally warranted, SAS-theory begins by acknowledging that these states are events to be explained. These events emerge in the relationship between *sets* of human requirements (denoted as a set, $R$) and available resources (denoted as a set, $A$). When empirically observed, both $R$ and $A$ are sets tied to a specific actor or group of actors. Consequently, in any given concrete situation or example, the sets $R$ and $A$ are explicitly indexed for the individual by $i$ in a group with 1,2…, to $n$ individuals, to define who possesses a specific requirement set $R_i$ and who is controlling a specific resource set $A_i$.

Set $A$ implies a given resource that has the capacity to satisfy human requirements. Although the SAS-theory definition of resource set $A$ is agonistic to its content, SRT's resource classes can be used to populate $A$. The six resource classes love, service, goods, money, information, status, differentiate a superset of $A$ containing these six classes (Foa & Foa, 1974, pp. 140–143):

$$A_{SRT} = \{A_{love}, A_{service}, A_{goods}, A_{money}, A_{information}, A_{status}\}$$

While $A$ can be any set, for simplicity, we let $A = A_{SRT}$, and henceforth use $A$ to signify $A_{SRT}$.

To evaluate if $A$ is scarce, sufficient or abundant, scholars must compare it to a set of *requirement*s $R$. $R$ can contain any defined requirement, including but not limited to needs and wants. By definition, SAS-theory is agnostic about the content of $R$ (Daoud, 2011a, 2018b). Consequently, $R$ captures an observable, empirical manifestation of underlying biological and social-psychological mechanisms. The manifestation might be any valid configuration of a requirement; anything a human utters, prefers indicates or selects. The nature of $R$ is, simply put, an empirical question. In some cases, $R$ might be defined as a range because $R$'s exact value would be imprecise to define categorically (e.g., the amount of water required to sustain life, or amount of time to complete a task).

Contrasting with SRT, 'needs' is the concept functioning most similar to $R$. Yet the difference between them implies more than a choice of words. SAS-theory relies on a layered ontology, differentiating the empirical (what is observed) from the trans-empirical (what is unobserved) (see Daoud, 2007). At the empirical level, researchers can only observe the manifestations of biological requirements that are mixed with cultural and psychological mechanisms. These observations manifest as (subjective) needs, wants, demands, and desires, crystallized as $R$. To be clear, such observations might include biomedical ones regarding biological requirements. At the trans-empirical level, however, SAS-theory allows for different conceptions of human nature and thus definitions of human qualities, such as a (normative) categorization of requirements as needs and wants. This separation between empirical and trans-empirical helps to avoid any normative conceptions that would prescribe any normative distinction of valid from invalid $\boldsymbol{R}$ (such as "false needs") in empirical work (Stillman, 1983). This layered ontology allows SAS-theory to conduct agnostic analyses of empirical data, and after that transgress the empirical level by introducing theories of non-observable qualities that might explain — or challenge — the empirical findings. Nonetheless, if we defer evaluating $\boldsymbol{R}$'s nature in terms of needs or non-needs, we can define the following SRT-compatible requirement set (Foa & Foa, 1974, pp. 140–143).

$$\boldsymbol{R}_{SRT} = \{\boldsymbol{R}_{love}, \boldsymbol{R}_{service}, \boldsymbol{R}_{goods}, \boldsymbol{R}_{money}, \boldsymbol{R}_{information}, \boldsymbol{R}_{status}\}$$

While $\boldsymbol{R}$ can take any elements, for simplicity, we let $\boldsymbol{R} = \boldsymbol{R}_{SRT}$ and henceforth in this chapter, we use $\boldsymbol{R}$ to signify $\boldsymbol{R}_{SRT}$.

All actors have requirements despite being satisfied or not, in contrast to "needs," which, at least in SRT, imply that they are only held by individuals lacking resource optimum, and thus actors holding resource optimum either have no needs. Alternatively, following Greenberg (1981), individuals experiencing sufficiency or abundance might have "felt needs," which are qualitatively different from (actual or objective) needs. Consequently, while the choice of words describing the empirical observation it denotes might include the word need, $\boldsymbol{R}$ is observed by asking people what they require or identifying indirect indications of requirements. $\boldsymbol{A}$ is measured by asking people what is available to them to satisfy their $\boldsymbol{R}$.

An individual possesses a set of resources, $\boldsymbol{A}$, that might or might not match her requirements, $\boldsymbol{R}$. The sets $\boldsymbol{R}$ and $\boldsymbol{A}$ are the two main concepts needed to distinguish SAS from each other. $\boldsymbol{R}$ and $\boldsymbol{A}$'s cardinality determines these events, that is, their relative sizes implied by the pipes | |. For example, the set $\boldsymbol{A} = \{\text{a house, a car}\}$ has the cardinality $|\boldsymbol{A}| = 2$, because this set has two objects. If the cardinality $|R|$ is larger than the cardinality $|A|$, it implies scarcity. This scarcity is implied by the relationship sign ">," $|\boldsymbol{R}| > |\boldsymbol{A}|$. When $|R|$ is smaller than $|A|$ implies abundance "<" and, when $|R|$ is equal to $|A|$ implies sufficiency captured by the relationship sign "=".

| | | |
|---|---|---|
| *Scarcity* | $|\boldsymbol{R}| > |\boldsymbol{A}|$ | The cardinality of the requirement set is strictly larger than the resources set. |
| *Abundance* | $|\boldsymbol{R}| < |\boldsymbol{A}|$ | The cardinality of the requirement set is strictly smaller than the resources set. |
| *Sufficiency* | $|\boldsymbol{R}| = |\boldsymbol{A}|$ | The cardinality of the requirement set is equal to the resources set. This approximation can either be defined exactly or as a range. |

An undefined relation is denoted with $\sim$ and can be used like the following, $\boldsymbol{R} \sim \boldsymbol{A}$. Such undefined relationships are useful when either the elements' cardinalities or qualities are yet to be evaluated. $\boldsymbol{R} \sim \boldsymbol{A}$ implies undefined SAS. Following SRT, we define the set of all relationships in the following way:

$$\boldsymbol{R} \sim \boldsymbol{A}$$
$$= \{R_{love} \sim A_{love}, R_{service} \sim A_{service}, R_{goods} \sim A_{goods}, R_{money} \sim A_{money}, R_{information} \sim A_{information}, R_{status} \sim A_{status}\}$$

In $\boldsymbol{R} \sim \boldsymbol{A}$, we assume the exchangeability of elements in the sets according to their usefulness. For example, for transportation, a "black horse" is as usable as a "white horse" or perhaps a "donkey" or a "camel." However, the quality of these creatures in a different context might differ significantly. For example, using a white horse for a wedding gift signifies a different quality of that resource to that of a donkey. The degree of exchangeability of resources in the same resources class is determined empirically.

How an individual dynamically shifts between SAS can be exemplified with a play from Shakespeare. His *Richard III* line "a horse, a horse, my kingdom for a horse!" implies that King Richard required a horse, $\boldsymbol{R} = \{horse\}$, and was willing to exchange his kingship for getting one, $\boldsymbol{A} = \{kingship\}$. To satisfy his requirements, King Richard III offers his kingship in exchange for an (available) horse. In the case of King Richard III, we can picture the two requirement classes: one for his means of transportation (i.e., the SRT-resource class *goods*) and one for his requirement of kingship (i.e., the SRT-resource class *status*), denoted by $\boldsymbol{R}_{goods} = \{horse\}$ and $\boldsymbol{R}_{status} = \{\emptyset\}$.

We can then formalize all the SAS relationships as follows:

$$\text{Scarcity} \quad |\boldsymbol{R}_{goods} = \{horse\}| > |\boldsymbol{A}_{goods} = \{\emptyset\}|$$

$$\text{Abundance} \quad |\boldsymbol{R}_{status} = \{\emptyset\}| < |\boldsymbol{A}_{status} = \{kingship\}|$$

If the proposed exchange is made (in the theater play, it is not), Richard would have a horse but no longer being a king, making both the relations $\boldsymbol{R}_{goods} \sim \boldsymbol{A}_{goods}$ and $\boldsymbol{R}_{status} \sim \boldsymbol{A}_{status}$ sufficient:

$$\text{Sufficiency} \quad |\boldsymbol{R}_{goods} = \{horse\}| = |\boldsymbol{A}_{goods} = \{horse\}|$$

$$|\boldsymbol{R}_{status} = \{\emptyset\}| = |\boldsymbol{A}_{status} = \{\emptyset\}|$$

A fundamental property of SAS-theory is that it is agnostic about the $\boldsymbol{R} \sim \boldsymbol{A}$ relation. By *agnostic*, we mean a theory that does not assign any scientific priority among the three SAS events. This agnosticism differs from, most distinctly, neoclassical economics, which assumes the normative position of limiting scientifically and socially relevant problems to those caused by scarcity, $|\boldsymbol{R}| > |\boldsymbol{A}|$ (e.g., Friedman, 2009; Zinam, 1982). This assumption forces any occurrence of abundance, $|\boldsymbol{R}| < |\boldsymbol{A}|$, to be explained as a symptom of some underlying scarcity (Zinam, 1982). Most conspicuously, John Maynard Keynes (2008) argued that an abundant supply is the consequence of scarce demand. A suggestion that might not always be true (see, e.g., Bååth, 2018; Daoud, 2007). As we have argued earlier in this chapter, a normative prioritization of any SAS as especially problematic (or preferrable) is unwarranted, because they may all exist and have consequences. Moreover, there are cases in which scarcity is the desired state over abundance. For example, creating scarce availability of — or even banishing — weapons of mass destruction is considered by many a noble ambition and desired goal. In other cases, people have been found to limit their requirements to achieve sufficiency, rather than increase available resources (Daoud, 2011b; Osikominu & Bocken, 2020).

Having laid out SAS and how they relate to each other, we turn to how SAS affects motivations for action. Assuming that SAS might influence human motivations to act, we suggest that there are several strategies that would be available to individuals experiencing a defined SAS event (Table 1).

*Table 1: A sketch of some individual and social strategies to cope with SAS (Daoud, 2018b)*

|  | Scarcity | Abundance | Sufficiency |
|---|---|---|---|
| Defensive (avoid it) | (1) Debt | (2) market efficiency | (3) Greed, gluttony |
| Reactive (reduce it) | (4) Protestant work ethic, simplifying, austerity | (5) homophily, stereotypes, | (6) opulence, self-destructive behavior |
| Adaptive (embrace it) | (7) Innovation | (8) Serialism, multiculturalism | (9) Modesty, frugality, environmentalism |
| Creative (inflate it) | (10) Sadism, Masochism, ritual sacrifice, speculation | (11) Lavishness, the sacred, feasting | (12) Generosity, charity. |

*Notes*: (a) Adopted from Abbott (2014) and augmented to cases of scarcity and sufficiency. (b) The concepts provided are indicatory and not exhaustive.

As shown in Table 1, the strategies laid out there suggests that individuals might act differently under a given resource state, generally adopting a coping strategy which includes one of four ambitions: to avoid, reduce, embrace, or inflate the resource state in question. These strategies contrast the motivations for exchange in SRT (Foa & Foa, 1974, p. 126). Rather than starting from intrinsic psychological motives, grounded in a trans-empirical conception of human nature, the strategies are derived from observable human action, and thus agnostic of human nature. The agnosticism implies that SAS-theory does not demand a definition of human nature to be employed on an empirical level. While it is beyond this chapter to identify all possible human action motivations, we will henceforth focus on motivations provoked by SAS. To do that, we incorporate a layered definition of SAS encompassing how an individual relates to another one and how an individual relates to a social system.

*On the layered systems of SAS*

The different strategies fostered by SAS events must be understood in the light of how the sets $\boldsymbol{R}$ and $\boldsymbol{A}$ are related, which they can be in two manners. Being grounded in critical realism, SAS-theory stresses the importance of distinguishing what level of reality an analysis regards (Daoud, 2011a; Sayer, 2000). All reality levels are open systems, which might be influenced by other layers, with one exception.[1] That exception is if the highest level of reality is defined as a closed system, implying that it suffers no extrinsic influence. We will limit the argument to the distinction between two levels: individual and systemic.

The distinction between the individual and the systemic level might be conceived as SAS-theory's "cultural psychology." The individual, indexed by *i*, is the level of experience where $\boldsymbol{R}_i$ and $\boldsymbol{A}_i$ operate. Methodologically, this means observing the sociopsychological accounts of individuals. Generally, SRT-studies tend to operate on this level, which will have reasons to come back to. The system level, in contrast, is the aggregate of individual-level experiences within an analytically closed system. Such a system might differ in size and complexity, ranging from small groups to the world's total population. Thus, the system includes SRT's institutions that define legitimate forms of exchange (Foa & Foa, 1974, pp. 150–152), but is not limited to these exact structural features. In this case, we define the system level as a closed system encompassing all individuals, including collective social phenomena such as institutions.

The system level, indexed by *s*, is formalized by aggregating individual $\boldsymbol{R}$ and $\boldsymbol{A}$, generating systemically available resources $\boldsymbol{A}_s$ and requirements $\boldsymbol{R}_s$. These two sets aggregates resources and requirements in the system in the following way: $\boldsymbol{A}_s = \bigcup_{i=1}^{n} \boldsymbol{A}_i$ denotes the union of all available resources in the system with *n* individuals, and $\boldsymbol{R}_s = \bigcup_{i=1}^{n} \boldsymbol{A}_i$ denotes the union of all requirements in the system.[2] $\boldsymbol{R}_s$ thus indicates aggregate observations of conceptions regarding material needs, social conventions, religious dogmas, biological conditions, political ideologies, norms, etc. that informs requirements for a given resource in a defined system. In effect, the distinction of and relation between $\boldsymbol{R}_i \sim \boldsymbol{A}_i$ and $\boldsymbol{R}_s \sim \boldsymbol{A}_s$ is the primary concern for SAS-theory, which, at a later stage, offers explanations of different observed strategies for coping with SAS (cf. Table 1).

So far, we have considered the direct relation of $\boldsymbol{R}$ and $\boldsymbol{A}$ (i.e., $\boldsymbol{R} \sim \boldsymbol{A}_-$), where no exchange is occurring. By differentiating individual and system level, it becomes possible to relate how elements in sets for the individual *i*, $\boldsymbol{R}_i \sim \boldsymbol{A}_i$, are exchanged with the elements in sets for another individual *j*, $R_j \sim A_j$, including their exchange with $R_s \sim A_s$ of the system *s*. We achieve this exchange via the concept entitlements (Daoud, 2010, 2018a; Reddy & Daoud, 2020).

---

[1] See Koumakhov and Daoud (Forthcoming, 2017) for how critical realism and social psychology overlap, and used for SRT.
[2] Mathematically, a regular set is defined by containing at maximum one element of the same type. We use a *multiset* (bags) definition that unlike a regular set accommodates several (multiple) instances for each element.

*Entitlements*, denoted *E*, are relationships between individuals, or individuals and system, defining the conditions for how they exchange resources they currently possess for resources they yet do not possess. This concept relates to Foa and Foa's conception of institutions (1974, pp. 150–152), producing various empirical effects of the acceptable conditions for exchange. Defined by culture and society, entitlements similarly rest on rules or conventions of legitimacy.

Entitlements are nested recursive relationships between individuals and systems. For example, I own a car. Why is this ownership accepted? This ownership is accepted because I exchange money to buy this car. Why is my ownership of money accepted? Because I got this money by exchanging my house for a profit. Why is my house ownership accepted? Because I inherited it from my mother. Why was her ownership accepted? And so on and so forth. Based on legitimacy, the chain of exchange recursively continues. Each link in this chain of entitlements rests on socially accepted rules for the exchange of resources.

We build on Amartya Sen's definition of entitlements:

> A person's ability to command food—indeed, to command any commodity he wishes to acquire or retain—depends on the entitlement relations that govern possession and use in that society. It depends on what he *owns*, what *exchange possibilities* are offered to him, what is *given to him free*, and what is *taken away* from him. (Sen, 1983, pp. 155–156, emphasis added)

In this quote, Sen defines entitlements as access to four types of relations that enable an individual to acquire resources: ownership, trade, gifts, and extraction. Entitlements are thus culturally or politically defined relations of fair exchange according to one of the four types. The four types of entitlements compare to Foa and Foa's five paradigms of exchange in SRT (1974, pp. 178–180). Moreover, while individuals might have different entitlements, an entitlement exists on a systemic level. That is the case, since, for anything to be socially legitimate, its legitimacy has to be generally accepted or at least tolerated. For example, citizenship in a welfare state grants entitlements legitimizing the exchange of various resources for the status of citizenship. In effect, entitlements might be restricted to specific individuals or groups in a certain society. The most prominent cause would be a feudal society, in which the estate of an individual would determine her entitlements, while the entitlements differed vastly between different estates.

Mathematically, the exchange of resources is then defined by the entitlements function $E_{i(\cdot)}(A_i, A_{(\cdot)})$ that maps an exchange of resources between individual *i* and another actor (either the system or an individual). For clarity, we define the other actor as the individual *j*. An individual does not exchange with him or herself, we assume $j \neq i$. We define $E_{ij}(A_i = a_{ik}, A_j = a_{jz})$ as the entitlement function's that transform two individuals' resource sets. This function and transformation is defined by whatever exchange conditions individual *i* and *j* agreed to be legitimate. $E_{ij}$ defines a relation between two sets of available resources $A_j$ and $A_i$, possessed by *j* and *i*. For example, *j* might be employed by *i*. By working for *i*, *j* gives her labor and thus is entitled to some of *i*'s available resources. $E_{ij}$ then functions similar to an exchange rate, defining how much or what of $A_i$ that is legitimately transferred to $A_j$ due to the employment-relation.

Expanding our formalizations of SAS, we add entitlements to the relation denoted as $R \sim E(A)$. To describe an individual's perspective exchanging with a system, we use the notation $R_i \sim E_{is}(A_i, A_s)$. This notation is useful when an individual exchanges with "all other individuals' *A*" in some generalized system such as the stock market or a welfare state. An entitlement denoting the relation between two individuals would work in the same way. On the system level, the relation would still be $R_s \sim A_s$ since the system itself does not have entitlements. The system cannot be entitled to anything that is not already available within the system, since nothing exists outside the system. Entitlements at the system level would only be relevant if two systems are exchanging with each other—for example, two countries engaged in international trade.

On the individual level, we might then evaluate which of SAS the individual experiences. Suppose that an individual *i* requires all six resource classes, the cardinal definition of $|R_i|$. To what extent these requirements are fulfilled is then defined by the sets of resources the entitlement function $E_{is}(A_i, A_s)$. makes available for the individual. Failure or success on the entitlement function determines which SAS event the individual experiences. The outcome depends on the entitlement's ability to fulfill the individual's requirements. To evaluate SAS, we remain agnostic about why the entitlement has succeeded or failed. To explain SAS, we evaluate the nature of *E*. An entitlement might fail both because its institutional design is flawed, thus failing to meet the individual's requirements while working as intended, and because one of the involved individuals breaks the rules, disavowing the entitlement's legitimacy. In the end, what the failure consists of would be an empirical question. We denote success as $E^+$ and failure as $E^-$. On the individual level, the difference between SAS would be as follows:

*Scarcity*  $\quad |R_i| > |E_{is}^-(A_i, A_s)|$

*Abundance*  $\quad |R_i| < |E_{is}^+(A_i, A_s)|$

*Sufficiency*  $\quad |R_i| = |E_{is}^+(A_i, A_s)|$

By introducing entitlements, we have a function for what resources are available to an individual through an exchange. To be clear, the pipes imply the cardinality of the result of the entitlement function, since the entitlement function cannot operate on a cardinal set. Effectively, if entitlements would change, the available resources for an individual would change. These are some of the new SRT questions scholars can embark on to investigate. Having defined SAS on an individual level, we now show how they produce new types of SAS.

*Quasi-SAS*

Individuals' experience of SAS varies depending on the system's aggregate resource state. Here, we relate individual sets of $R_i \sim E_{is}(A_i, A_s)$ *to the* $R_s \sim A_{s-}$ set of the system in which the individual exists. To enable such comparisons is a core rationale of SAS-theory and the advantage of distinguishing reality layers. By evaluating individual SAS in relation to the systemic SAS, we distinguish between if an individual experiences an absolute- or quasi-SAS. While absolute- and quasi-sufficiency is possible, we focus on events of absolute- and quasi-scarcity and abundance.

We distinguish absolute scarcity from quasi-scarcity as follows. *Absolute scarcity* is when scarcity on the individual level coexists with scarcity on the systemic level. The individual's entitlement fails to make the required resources available, simply because there are not enough to go around.

|  | *Individual* | *System* |
|---|---|---|
| Absolute scarcity: | $|R_i| > |E_{is}^-(A_i, A_s)|$ | $|R_s| > |A_s|$ |

However, individuals might also experience scarcity even though the required resource is not scarce on a systemic level, but sufficient or abundant. Consequently, the individual's entitlement function fails to make the required resource available, but the failure is caused by something different than absolute scarcity. This state is what we call *quasi-scarcity*.

|  | *Individual* | *System* |
|---|---|---|
| Quasi-scarcity: | $|R_i| > |E_{is}^-(A_i, A_s)|$ | $|R_s| \leq |A_s|$ |

For abundance, the relation between individual and system is the mirror image of scarcity. *Absolute abundance* is when the individual experiences an abundance of a required resource, which is also abundant on the systemic level. In this case, the entitlement function makes more resources available than requires.

|  | *Individual* | *System* |
|---|---|---|
| Absolute abundance: | $|R_i| < |E_{is}^+(A_i, A_s)|$ | $|R_s| < |A_s|$ |

*Quasi-abundance* is then when the individual's entitlement function enables more resources than required, while the relevant resource is scarce or sufficient on a systemic level.

|              | *Individual*                          | *System*              |
|--------------|---------------------------------------|-----------------------|
| Quasi-abundance: | $\|\boldsymbol{R}_i\| < \|E_{is}^{+}(\boldsymbol{A}_i, \boldsymbol{A}_s)\|$ | $\|\boldsymbol{R}_s\| \geq \|\boldsymbol{A}_s\|$ |

The ability to distinguish between absolute and quasi-states of SAS is useful because they warrant different descriptions on the empirical level and different explanations on the trans-empirical level. On the empirical level, the "quasi" prefix implies that the satisfaction of $\boldsymbol{R}_i$ deviates from the satisfaction of $\boldsymbol{R}_s$. In effect, an individual experiencing a quasi-version of a certain SAS-state might be motivated to act quite differently from one experiencing an absolute version of the same SAS-state. That is the case because quasi-states imply inequality, in its most descriptive sense.

Enabled by the definition of the entitlement function, these distinctions between absolute and quasi-SAS tie in with the core ambition of SRT: to identify how mechanisms of resource exchange are characterized among social groups and their normative implications. The concept entitlement enables explanations of how the social group (here, system) influences the individual's possibility to, and motivations for, exchange. Foa and Foa define five paradigms of interaction that might occur in exchange. These paradigms are different combinations of two actors exchanging, where the first proactively gives or takes, and the second reacts by giving or taking (Foa & Foa, 1974, p. 179). By distinguishing the individual from the systemic level and defining the entitlement function that governs exchange, SAS offers a means of explaining how individual and systemic SAS-states might cause a given exchange, and what kind of entitlements are in place when a particular exchange is carried out. Entitlements link individual and systemic levels, and thus provides SRT with a theoretically more developed explanation of the conditions and constraints that might influence resource exchange, both in general and to specific resource types.[3]

The distinction between absolute and quasi-SAS has implications for how resource states influence motivations for exchange. For example, an individual experiencing quasi-SAS might be morally engaged because he or she does not have enough resources despite that the system has abundant resources. Those experiencing quasi-abundance might resist any change to the entitlement system that jeopardizes the current unequal distribution of resources. In contrast, an individual experiencing an absolute abundance might be content with the situation if the resource is of general value (e.g., food). From an explanatory perspective, quasi-SAS might have arisen from inequality, kleptocracy, clientelism, and similar. Events that historically has fostered social transformations ranging from violent revolts to the institutionalization of black markets.

---

[3] Sen offers a different formalization of exchange entitlements based on social welfare theory (Sen, 1983, Appendix A).

**How SRT benefits from SAS-Theory: Two Applications**

Our analytical distinctions of absolute- and quasi-SAS allow SRT to benefit in at least two ways: (A) an agnostic approach to motivations for exchange and (B) a social systems approach to SAS. By conceptualizing SRT through SAS-theory, we can bring abundance and sufficiency into SRT. We demonstrate these benefits by applying SAS-theory to two examples: Max Weber's protestant's ethic-turn-capitalist spirit and global hunger in a world of food abundance.

*Application I: The protestant ethic and abundance-based motivations*

One of the most distinct consequences of the assumption of scarcity in SRT is that it limits the possible motivations for an actor who holds an abundance of resources. An actor experiencing abundance, we argue, might use one of four different strategies: Defensive (avoid it), Reactive (reduce it), Adaptive (embrace it), or Creative (inflate it) (Daoud, 2018b; see also Abbott, 2014). Foa and Foa suggest two sources of abundance-based motivation: "pain and discomfort" and "satiation" (Foa & Foa, 1974, pp. 128, 138). These two motivations would warrant the actor to use a reactive strategy, or possibly a defensive one. In addition, Greenberg (1981) suggests that actors may experience "felt needs", despite not actually possessing scarce resources, which would motivate certain actors to use creative or adaptive strategies. However, SAS-theory uses requirements which includes any felt needs, wants, or preferences. From our SRT-SAS synthesis, requirements are the empirical manifestations of human needs mixed with psychological and cultural expressions.

People who knowingly possess an abundance of resources still tend to use creative and adaptive strategies. Weber's (2005) study of frugal protestants-turning-capitalists offers an informative example. The protestant ethic's emphasis on saving and investing excess capital, rather than spending or destroying it, can be interpreted as a strategy for reducing abundance to scarcity. Because Foa and Foa suggest that money might have an infinite optimal range (1974, pp. 125–128), we simplify our example by focusing on capital as the resource class "goods." The protestant, indexed by *p*, has the requirement $\boldsymbol{R}_p$ and the available resources $\boldsymbol{A}_p$, which's cardinalities stand in the following relation:

$$\left|\boldsymbol{R}_{p_{goods}} = \{porcelain, copper\}\right| < \left|\boldsymbol{A}_{p_{goods}} = \{porcelain, copper, silk\}\right|$$

The protestant thus possesses an abundance of goods. However, by striking a deal with a merchant, indexed as *m*, the protestant is entitled to investing some his silk in the merchant's training company for a promise of a future return of a greater quantity of silk when the merchant has increased his shares of the silk market. The entitlement is denoted $E_{pm}$. Assuming that the entitlement exchange is successful between the protestant and merchant, the relation now looks as follows:

$$\left|\boldsymbol{R}_{p_{goods}} = \{porcelain, copper\}\right| = \left|\boldsymbol{A}_{p_{service}}^{E_{pm}^+(\boldsymbol{A}_{p_{goods}} = \{porcelain, copper\},} = \{a\ promise\ of\ future\ silk\}, \boldsymbol{A}_{m_{goods}} = \{silk\})\right|$$

By investing the silk, the merchant has achieved sufficiency. While he has been given a promise in exchange, the promise is not classified as a good, but as a service which might later provide goods, so far that the promise is honored. While the protestant is awarded something in exchange for his silk, that is not good. The relation between requirements and available resources for goods is that of sufficiency. However, this exchange was motivated by neither a felt need for silk (as silk is abundant), nor pain and discomfort from satiation, as the protestant could then have destroyed or given the silk away freely. However, by relating the protestant's requirements and resources to the system level, we find justifications for assuming specific motivations. We evaluate first system abundance and then system scarcity of silk.

If silk is abundant on the system level, it means that the aggregate requirement $R_s$ for silk overshoots the silk available within the system $A_s$:

$$\left|R_{s_{goods}} = \{silk\}\right| < \left|A_{s_{goods}} = \{silk\}\right|$$

In this case, the protestant possesses an absolute abundance of silk. Ridding himself of the abundant silk, even if it is only exchanged for the promise of a future, the greater quantity of silk, would be motivated by the pain and discomfort of risking that the silk would never be utilized for any meaningful end, such as profits or beautiful clothing. Foa and Foa's suggestion that super optimum would self-regulate to optimum because of the troubles caused by the abundance may be true. However, the protestant might as well be motivated by the experience of abundance to employ a reductive strategy, including but not limited to pain and discomfort. Any reason for attaining sufficiency would motivate such a strategy, not only negative stimuli but also by generosity. While the SRT definition of motivations limits the possible motivations to one, the SAS-theory's agnostic approach enables the inclusion of motivations stemming from other types of requirements.

If silk is scarce on the system level, it means that the aggregate requirement for silk overshoots the silk available within the system:

$$\left|R_{s_{goods}} = \{silk\}\right| > \left|A_{s_{goods}} = \{silk\}\right|$$

In this case, the protestant possesses a quasi-abundance of silk. While silk is required in the system, the protestant himself does not require silk. The motivation for investing the quasi-abundant silk for a promise of a future, greater abundance of silk, cannot be rooted in pain or discomfort as in the previous case. Nor does the exchange fulfill any alternative need. Consequently, the protestant's motivation is unexplainable in terms of scarcity or needs. However, the protestant's motivation is explainable if we consider the investment a creative strategy for dealing with abundance. In other words, the protestant does not require the silk now, but he might require a greater quantity in the future. Such a strategy might be motivated by possessing quasi-abundance because the aggregate demand means that it would be a good opportunity to invest the silk to (hopefully) acquire a greater quantity of silk in a future of unknown SAS. In this case, the protestant's strategy relates to SRT's concept "power" (Foa & Foa, 1974, pp. 134–135). The protestant might give the abundant silk away out of charity, or, as in this case, invest it for (possible) future profits.

*Application II: Famine in an abundant global food system*

As previously mentioned, the global food supply is absolutely abundant and has so been since the UN started to measure it in the 1960s. Currently, the global food supply overshoots the basic human caloric need by approx. 60% (FAO et al., 2001, 2015). Despite this abundance, 10% of the global population suffers from famine. This is a case of quasi-scarcity.

In this case, we thus define $R_i$ as the human caloric requirement, labeled "food." Mathematically, the following is true for about one billion of poor individuals that are starving every day:

*Individual*

$$\left|R_{i\_goods} = \{food\}\right| > \left|A_{i\_goods} = \{\emptyset\}\right|$$

*System*

$$\left|R_{s\_goods} = \{food\}\right| > \left|A_{s\_goods} = \{food\}\right|$$

This state of food quasi-scarcity begs the question — are all people entitled to food? According to article 25 of the UN's universal declaration of human rights, "Everyone has the right to a standard of living adequate for the health and well-being of himself and of his family, including food…" (The Universal Declaration of Human Rights, 1948, n.p.). This article states that every individual, indexed by *i*, is entitled to sufficient food to avoid starvation and humanity's degradation. The entitlement charges the system, indexed by *s*, with providing the food. The exact mechanism through which the food is supplied may vary a lot between individuals and does not have to be defined. Article 25 states that everybody is entitled to some functional means of acquiring food, might that be buying it on a capitalist market or hunting and gathering. However, this entitlement function fails to make enough food available through any available mechanism for all starving people.

*Individual*

$$\left|R_{i\_goods} = \{food\}\right| > \left|E_{is}^{-}(A_{i_{goods}} = \{\emptyset\}, A_{s_{goods}} = \{food\})\right|$$

*System*

$$\left|R_{s_{goods}} = \{food\}\right| < \left|A_{s_{goods}} = \{food\}\right|$$

In the equation, the starving individual's entitlement fails, resulting in that the set of available resources for the individual turns out empty. This failure does not tell us what distribution mechanisms that the individual hypothetically could have used to succeed. Any kind of mechanism, from capitalist markets to foraging, fails in this scenario. For an individual who does not starve, and the entitlement succeeds to supply food, the following is true:

*Individual*

$$\left|R_{i_{goods}} = \{food\}\right| \leq \left|E_{is}^{+}(A_{i_{goods}} = \{food\}, A_{s_{goods}} = \{food\})\right|$$

*System*

$$\left|R_{s_{goods}} = \{food\}\right| < \left|A_{s_{goods}} = \{food\}\right|$$

The non-starving individual's entitlement succeeds in the equation, making enough food available through some viable mechanism. As follows on Article 25, all humans have a right not to starve, but the mechanism which provides the food might differ as long as non-starvation is the outcome. In this case, individual experiences sufficiency or scarcity, begging the question of why the entitlement succeeds for some people but not others.

The IMF is one of the most influential institutions in the trade of improving global economic conditions. Reviewing their demands on food-supply for lender states, the IMF generally demands liberalized markets as a means of improving the (economic) efficiency of food supply (Daoud et al., 2019). As shown in Table 1, market liberalization is one commonly suggested defensive strategy for coping with abundance. Liberalized markets are often assumed to maximize efficiency; minimize the risk of unrequired abundances while minimizing scarcity simultaneously through the signal-system of prices (Friedman, 2009; Hayek, 1945). However, in practice, liberalized markets do not seem to have that effect. Instead, the already wealthy are over-supplied (Daoud, 2007), and the practices of pricing reproduce taken-for-granted social structures (Bååth, Forthcoming). An increasing number of studies find that the huge trade surpluses of the world's wealthiest nations actually obstruct poorer nations' development of local systems of provisioning, foremost enriching the already entitled while not redeeming famine among the poor (see McGoey, 2018). In the US, for example, the poorest neighborhoods now have the least food outlets while also being subject to the steepest food prices due to the absence of competition and large-scale supermarkets (Walker et al., 2010). To conclude, while liberalized markets might be an efficient strategy for reducing the risk of food scarcity on a system level, it does not seem to deliver equally well on an individual level. That is the case; it seems because liberalized markets make wealthy consumers end up with an abundance of food, before feeding the poor consumers. Instead of minimizing scarcity on the individual level, it inflates or at least adapts the abundance of a sub-set of wealthy individuals.

Considering wealth as an entitlement *E*, rather than a resource, enables us to describe this case of quasi-scarcity better, establishing the exchange rate of food between the individual, indexed by *i*, and the global food system, indexed by *g*. By defining wealth as an entitlement, it suggests that wealthy individuals possess more money to buy food and have access to more and better offers of food. In effect, it is of less importance what precisely the individual has to exchange for food. Whatever she has, her failing entitlement makes insufficient amounts of food available in return, despite there being an abundance of food in the world.

*Individual*

$$\left|R_{i_{goods}} = \{food\}\right| > \left|E_{ig}^{-}(A_{i_{goods}} = \{\emptyset\}, A_{g_{goods}} = \{food\})\right|$$

*System*

$$\left|R_{s_{goods}} = \{food\}\right| < \left|A_{g_{goods}} = \{food\}\right|$$

In the equation, the entitlement's failure implies that the individual's lack of wealth means that enough food to cover the human caloric requirement is not made available by any means. This might be due to a shortage of cash, but equally well the absence of food vendors. Additionally, it might imply the financial inability to stave off land-grabbers, losing land access for sustenance farming or hunting and gathering. In the end, all these scenarios have the same outcome: starvation due to a lack of wealth. However, the wealth-function of the individual who does not starve does deliver sufficient or abundant quantities of food.

*Individual*

$$\left|R_{i_{goods}} = \{food\}\right| \leq \left|E_{ig}^{+}(A_{i_{goods}} = \{food\}, A_{g_{goods}} = \{food\})\right|$$

*System*

$$\left|R_{g_{goods}} = \{food\}\right| < \left|A_{g_{goods}} = \{food\}\right|$$

These relations tell us that the entitlement of wealth is dramatically unevenly distributed among the world's population. Following the entitlement function, the global food system, supplying food through liberalized markets, is an efficient strategy for exchanging food (goods) for wealth (money). However, some individuals hold significantly more wealth than others. As a result, the current global food system's orientation towards wealth entitles the wealthiest more food than they require before meeting all individual requirements (see also Sen, 1983, pp. 154–157). Moreover, an actor in a state of quasi-scarcity of food would be much more prone to revolts or riots. Food riots under quasi-abundance have the possible outcome of actually feeding the rioting hungry, reducing famine, while food riots under absolute abundance may either be futile or possibly redistribute some quasi-sufficiency or abundance, while not reducing famine.

Moreover, the experience of quasi-scarcity implies that another individual's abundance might very well serve as a provocation that motivates riots or other kinds of social upheaval, and legitimizes violence as an entitlement. To avoid such upheaval, policymakers implement social policies such as minimum income or food coupons to give temporary entitlements for the poor (Conklin et al., 2019;

Nandy et al., 2016). In the opposite case, foods have been systemically destroyed to mitigate abundance, effectively protecting wealth as an entitlement (Prasad, 2012).

**Discussion and Conclusions**
This chapter synthesized SRT and SAS, enabling SRT to analyze abundance and sufficiency in tandem with scarcity for analyzing resource exchange. First, we outlined how SRT rests on an assumption of scarcity as the primary resource state causing exchange motivations and the problems caused by that assumption. Second, we used SAS-theory to formalize SAS in an agnostic manner, tying them to different behavioral strategies that individuals use when engaging with a specific resource state. Third, we formalized the relation between individual and systemic level SAS. This relation is influenced by entitlement functions, allowing the distinction between an individual's experiencing absolute and quasi-SAS. This difference is essential, as quasi-SAS implies different exchange motivations and strategies than absolute SAS. Lastly, we exemplified how the formalized relations through two examples show how abundance-based motivations function, and how quasi-scarcity requires different explanations than absolute scarcity.

The rationale of this chapter rested on the undertheorizing of two aspects of SRT. The first one regards how abundance and sufficiency affect exchange. By formalizing SAS's relations, SRT has a precise language to identify how sets of resources and requirements change through an exchange. By doing so, SRT achieves an agnostic outset for analyzing resource exchange and its underlying motivations, while still acknowledging existing findings regarding the effects of needs in exchange. An important consequence of using this definition is that it would enable SRT to redefine how abundance relate to resources exchange. While SRT defines negative abundances, which fosters the arousal of new needs (Foa & Foa, 1974, pp. 128, 138; Greenberg, 1981), the agnostic approach enables SRT to define also events of positive abundance.

The second aspect regards the undertheorizing of the relation between different social reality levels, primarily between the individual and the (societal) system levels. In general, SRT focuses on the individual level, which tends to leave out macro-scale social, political, and cultural processes that will affect an individual's exchange behavior. By distinguishing the individual level from the systemic level, SRT might consider different individuals' entitlements and how they influence the individual's resource state. Thereby, SRT is enabled to identify and analyze both absolute and quasi-SAS. Such identification and analyses provide more in-depth knowledge of an individual's motivations to engage in exchange in a particular manner.

Our proposal relates to Greenberg (1981), suggesting that macro processes influence requirements' formation, including needs. Greenberg's critique primarily regards that participants in exchange experiments might be affected by their everyday experience and understanding of need. However, it is equally applicable to the scientific study and theories of exchange in general. Our argument shows that SRT mechanisms exist on different levels of reality and that they might offer better explanations of exchange situations by explicating how they function on these levels.

Synthesizing SAS and SRT opens up new and deepened predictions, hypotheses, and explanations of exchange dynamics. For example, a recent re-examination of the "marshmallow test" suggests how such developments might offer essential insights (Watts, Duncan, & Quan, 2018). This test consists of an exchange situation. While the test has different versions, the main version consists of young children offered a marshmallow, promising an additional one if they successfully abstain from eating the first one for some time. Children's performance in this test correlates with their performance as adults. The marshmallow tests' findings have shown that children with the psychological ability to delay gratitude and wait until they are awarded the second marshmallow achieve greater social and economic success later in life. This finding has led to the conclusion that there is an (unidentified) causal link between the individual's ability to delay gratitude and life-course success. However, a recent study by the team who first established the marshmallow test now finds that the supposed causal link between gratification delay and social success is rendered insignificant when controlling for the participants' socioeconomic status (SES) (Watts, Duncan, & Quan, 2018). Exactly how SES relates to gratification delay is yet to be explained, but it does imply that the children's everyday SES-based entitlements, and experiences of entitlement failure or success, play into how they engage with the test. Using our synthesis of SRT and SAS-theory enables further sophisticated analyses of the relation between individual and systemic scarcity, entitlements, and different resource classes' exchange dynamics. The synthesis might, therefore, offer new insights into the social psychology of SES. Conversely, SAS-theory might develop more sophisticated approaches to less concrete and more personal resources, such as love and status, by engaging with SAS.

Throughout this chapter, we have stressed the agnosticism of SAS-theory, emphasizing its descriptive and analytical usefulness in resource exchange studies. However, this agnosticism does not mean that normativity does not have a place in SRT. While SAS-theory's main ambition is to offer useful descriptions, the results can still be interpreted in a normative framework. For example, SRT is an essential theory in studies of distributive justice. By the variants of quasi- and absolute SAS, our synthesis enables distributive justice studies to explicate the relation between systemic conditions for just exchanges and the individual's engagement in such exchanges.

**References**

Abbott, A. (2014). The Problem of Excess. *Sociological Theory*, *32*(1), 1–26. https://doi.org/10.1177/0735275114523419

Archer, M. S., & Tritter, J. Q. (2000). *Rational Choice Theory: Resisting Colonization*. Psychology Press.

Bååth, J. (Forthcoming). *Towards a unified definition of price: Turning to pricing in practice*.

Bååth, J. (2018). *Production in a State of Abundance: Valuation and Practice in the Swedish Meat Supply Chain* [Doctoral thesis]. Uppsala University.

Baumgärtner, S., Becker, C., Faber, M., & Manstetten, R. (2006). Relative and absolute scarcity of nature. Assessing the roles of economics and ecology for biodiversity conservation. *Ecological Economics*, *59*(4), 487–498. https://doi.org/10.1016/j.ecolecon.2005.11.012

Becker, G. S. (1965). A Theory of the Allocation of Time. *The Economic Journal*, *75*(299), 493–517. JSTOR. https://doi.org/10.2307/2228949

Boudon, R. (2003). Beyond Rational Choice Theory. *Annual Review of Sociology*, *29*(1), 1–21. https://doi.org/10.1146/annurev.soc.29.010202.100213

Bronfenbrenner, M. (1962). The Scarcity Hypothesis in Modern Economics. *The American Journal of Economics and Sociology*, *21*(3), 265–270.

Conklin, A. I., Daoud, A., Shimkhada, R., & Ponce, N. A. (2019). The impact of rising food prices on obesity in women: A longitudinal analysis of 31 low-income and middle-income countries from 2000 to 2014. *International Journal of Obesity*, *43*(4), 774–781. https://doi.org/10.1038/s41366-018-0178-y

Daoud, A. (2007). (Quasi)Scarcity and Global Hunger. *Journal of Critical Realism*, *6*(2), 199–225. https://doi.org/10.1558/jocr.v6i2.199

Daoud, A. (2010). Robbins and Malthus on Scarcity, Abundance, and Sufficiency. *American Journal of Economics and Sociology*, *69*(4), 1206–1229. https://doi.org/10.1111/j.1536-7150.2010.00741.x

Daoud, A. (2011a). *Scarcity, Abundance and Sufficiency: Contributions to Social and Economic Theory* [Doctoral thesis]. Gothenburg University.

Daoud, A. (2011b). The Modus Vivendi of Material Simplicity: Counteracting Scarcity via the Deflation of Wants. *Review of Social Economy*, *69*(3), 275–305. https://doi.org/10.1080/00346764.2010.502832

Daoud, A. (2018a). Synthesizing the Malthusian and Senian approaches on scarcity: A realist account. *Cambridge Journal of Economics*, *42*(2), 453–476. https://doi.org/10.1093/cje/bew071

Daoud, A. (2018b). Unifying Studies of Scarcity, Abundance, and Sufficiency. *Ecological Economics*, *147*, 208–217. https://doi.org/10.1016/j.ecolecon.2018.01.019

Daoud, A., Reinsberg, B., Kentikelenis, A. E., Stubbs, T. H., & King, L. P. (2019). The International Monetary Fund's interventions in food and agriculture: An analysis of loans and conditions. *Food Policy*, *83*, 204–218. https://doi.org/10.1016/j.foodpol.2019.01.005

Dugger, W. M., & Peach, J. T. (2009). *Economic Abundance: An Introduction* (1 edition). Routledge.

Elster, J. (2015). *Explaining Social Behavior: More Nuts and Bolts for the Social Sciences* (2nd ed.). Cambridge University Press. https://doi.org/10.1017/CBO9781107763111

FAO, IFAD, & WFP. (2015). *The State of Food Insecurity in the World 2015. Meeting the 2015 international hunger targets: Taking stock of uneven progress*. Food and Agriculture Organization of the United Nations.

FAO, WHO, & UNU. (2001). *Human energy requirements* (No. 1; Food and Nutrition Technical Reports). Food and Agriculture Organization of the United Nations. http://www.fao.org/docrep/007/y5686e/y5686e04.htm

Foa, U. G., & Foa, E. B. (1974). *Societal structures of the mind*. Thomas Books.

Friedman, M. (2009). *Price Theory*. The Richest Man in Babylon.

Greenberg, J. (1981). The Justice of Distributing Scarce and Abundant Resources. In M. J. Lerner & S. C. Lerner (Eds.), *The Justice Motive in Social Behavior: Adapting to Times of Scarcity and Change* (pp. 289–316). Springer US. https://doi.org/10.1007/978-1-4899-0429-4_13

Hayek, F. A. (1945). The Use of Knowledge in Society. *American Economic Review*, *XXXV*(4), 519–530.

Hedström, P., & Stern, L. (2008). Rational choice and sociology. In S. N. Durlauf & L. E. Blume (Eds.), *The New Palgrave Dictionary of Economics* (2nd Edition, pp. 872–877). Palgrave Macmillan.

Kahneman, D. (2011). *Thinking, Fast and Slow*. Farrar, Straus and Giroux.

Keynes, J. M. (2008). Economic Possibilities for our Grandchildren (1930). In L. Pecchi & G. Piga (Eds.), *Revisiting Keynes* (pp. 17–26). The MIT Press. https://doi.org/10.7551/mitpress/9780262162494.003.0002

Koumakhov, R., & Daoud, A. (Forthcoming). Decisions and Structures: Dialogue between Herbert Simon and Critical Realists. *British Journal of Management*.

Koumakhov, R., & Daoud, A. (2017). Routine and reflexivity: Simonian cognitivism vs practice approach. *Industrial and Corporate Change*, *26*(4), 727–743. https://doi.org/10.1093/icc/dtw048

McGoey, L. (2018). Bataille and the Sociology of Abundance: Reassessing Gifts, Debt and Economic Excess. *Theory, Culture & Society*, *35*(4–5), 69–91. https://doi.org/10.1177/0263276416637905

Menger, C. (2007). *Principles of Economics*. Ludwig von Mises Institute.

Mullainathan, S., & Shafir, E. (2013). *Scarcity: Why Having Too Little Means So Much*. Macmillan.

Nandy, S., Daoud, A., & Gordon, D. (2016). Examining the changing profile of undernutrition in the context of food price rises and greater inequality. *Social Science & Medicine*, *149*, 153–163. https://doi.org/10.1016/j.socscimed.2015.11.036

Osikominu, J., & Bocken, N. (2020). A Voluntary Simplicity Lifestyle: Values, Adoption, Practices and Effects. *Sustainability*, *12*(5), 1903. https://doi.org/10.3390/su12051903

Panayotakis, C. (2011). *Remaking Scarcity: From Capitalist Inefficiency to Economic Democracy*. Pluto Press.

Prasad, M. (2012). *The Land of Too Much: American Abundance and the Paradox of Poverty*. Harvard University Press.

Reddy, S. G., & Daoud, A. (2020). Entitlements and Capabilities. In E. Chiappero-Martinetti, S. Osmani, & M. Qizilbash (Eds.), *The Cambridge Handbook of the Capability Approach* (pp. x–x). Cambridge University Press.

Robbins, L. (2007). *An Essay on the Nature and Significance of Economic Science*. Ludwig von Mises Institute.

Sayer, A. (2000). Key features of critical realism in practice: A brief outline. In A. Sayer (Ed.), *Realism and social science* (pp. 10–28). SAGE Publications Ltd.

Sen, A. (1983). *Poverty and Famines*. Oxford University Press.

Springborg, P. (1981). *The problem of human needs and the critique of civilisation*. Allen & Unwin.

Stillman, P. G. (1983). Scarcity, Sufficiency, and Abundance: Hegel and Marx on Material Needs and Satisfactions. *International Political Science Review / Revue Internationale de Science Politique*, *4*(3), 295–310. JSTOR.

The Universal Declaration of Human Rights, (1948). http://www.ohchr.org/EN/UDHR/Documents/UDHR_Translations/eng.pdf

Törnblom, K. Y., & Nilsson, B. O. (2008). Effect of Matching Resources to Source on their Percieved Importance and Sufficiency. In U. G. Foa, J. Converse, K. Y. Törnblom, & E. B. Foa (Eds.), *Resource Theory: Explorations and Applications* (pp. 81–96). Emerald.

Törnblom, K. Y., Stern, P., Pirak, K., Pudas, A., & Törnlund, E. (2008). Type of Resource and Choice of Comparison Target. In U. G. Foa, J. Converse, K. Y. Törnblom, & E. B. Foa (Eds.), *Resource Theory: Explorations and Applications* (pp. 81–96). Emerald.

Turner, B. S., & Rojek, C. (2001). *Society and Culture: Scarcity and Solidarity*. SAGE.

Walker, R. E., Keane, C. R., & Burke, J. G. (2010). Disparities and access to healthy food in the United States: A review of food deserts literature. *Health & Place*, *16*(5), 876–884. https://doi.org/10.1016/j.healthplace.2010.04.013

Weber, M. (2005). *The Protestant Ethic and the Spirit of Capitalism*. Routledge.

Xenos, N. (2017). *Scarcity and Modernity*. Routledge.

Zinam, O. (1982). The Myth of Absolute Abundance: Economic Development as a Shift in Relative Scarcities. *American Journal of Economics and Sociology*, *41*(1), 61–76. https://doi.org/10.1111/j.1536-7150.1982.tb01668.x